# The Magellan Adaptive Secondary VisAO Camera: Diffraction-Limited Broadband Visible Imaging and 20mas Fiber Array IFS

Derek Kopon[*a], Laird M. Close[a], Jared Males[a], Victor Gasho[a], Katherine Follette[a]
[a]CAAO, Steward Observatory, University of Arizona, Tucson AZ USA 85721

## ABSTRACT

The Magellan Adaptive Secondary AO system, scheduled for first light in the fall of 2011, will be able to simultaneously perform diffraction limited AO science in both the mid-IR, using the BLINC/MIRAC4 10μm camera, and in the visible using our novel VisAO camera. The VisAO camera will be able to operate as either an imager, using a CCD47 with 8.5 mas pixels, or as an IFS, using a custom fiber array at the focal plane with 20 mas elements in its highest resolution mode. In imaging mode, the VisAO camera will have a full suite of filters, coronagraphic focal plane occulting spots, and SDI prism/filters. The imaging mode should provide ~20% mean Strehl diffraction-limited images over the band 0.5-1.0 μm. In IFS mode, the VisAO instrument will provide R~1,800 spectra over the band 0.6-1.05 μm. Our unprecedented 20 mas spatially resolved visible spectra would be the highest spatial resolution achieved to date, either from the ground or in space. We also present lab results from our recently fabricated advanced triplet Atmospheric Dispersion Corrector (ADC) and the design of our novel wide-field acquisition and active optics lens. The advanced ADC is designed to perform 58% better than conventional doublet ADCs and is one of the enabling technologies that will allow us to achieve broadband (0.5-1.0μm) diffraction limited imaging and wavefront sensing in the visible.

## 1. INTRODUCTION: MAGELLAN VISIBLE ADAPTIVE OPTICS, IMAGING AND IFS

The Magellan Clay telescope is a 6.5m Gregorian telescope located at Las Campanas Observatory (LCO) in Chile. The Gregorian design allows for a concave F/16 adaptive secondary mirror (ASM) that can be tested off-sky with a retro-reflecting optic at the fast (F/1) ellipsoidal conjugate (see Figure 1). With our partners and subcontractors, we have fabricated an 85 cm diameter adaptive secondary that uses 585 actuators with <1 msec response times and will allow us to perform low emissivity AO science. We will achieve very high Strehls (~98%) in the Mid-IR AO (8-26 microns) that will allow the first "super-resolution" and nulling Mid-IR studies of dusty southern objects. Our high order (585 mode) pyramid wavefront sensor (WFS) built by the Osservatoria Astrofisico di Arcetri (Esposito et al. 2008) is similar to that used in the Large Binocular Telescope AO systems. (For more on the Magellan ASM and mid-IR AO, see Close et al. these proceedings)

The ASM is identical in optical prescription to the LBT ASMs and will use all the same hardware and control software (for more on the LBT AO system, see Riccardi et al. 2008). The primary infrared science camera, BLINC/MIRAC4 (Hinz et al. 2009), will receive IR light from a dichroic beam splitter. Visible light reflected by the dichroic will be sent to an optical board (hereafter called the "W-Unit") that holds both the pyramid WFS and the VisAO camera/IFS. The VisAO imaging mode is designed to work from 0.5-1.0 μm and the IFS from 0.6-1.05 μm. The layout of the W-unit is shown in Figure 2.

The high actuator count of our ASM will allow us to obtain modest Strehls (~20%) in the visible (0.5-1.0 μm). Because the Vis AO camera is integrated into the WFS stage, we can select a beam splitter to steer a percentage of the WFS visible light into the Vis AO camera with 8.5 mas pixels. This capability allows us flexibility in choosing how much light to send to either the VisAO camera or the WFS, depending on guide star magnitude, seeing conditions, and science goals, while simultaneously observing in the IR with BLINC/MIRAC4. The visible science light can either go to the CCD47 for imaging or can be directed to a fiber bundle that has both coarse and fine plate scales for IFS spatially resolved spectroscopy.

The Magellan site provides excellent seeing conditions with $r_o$ frequently as high as 20 cm at 0.55 μm. Because of this, we expect that at λ~ 0.9 μm there will be AO correction on bright stars and that on good nights, moderate Strehls will be possible in the I and z bands. The resulting angular resolution will be a spectacular 20-30 mas (although the corrected

---
[*] dkopon@as.arizona.edu

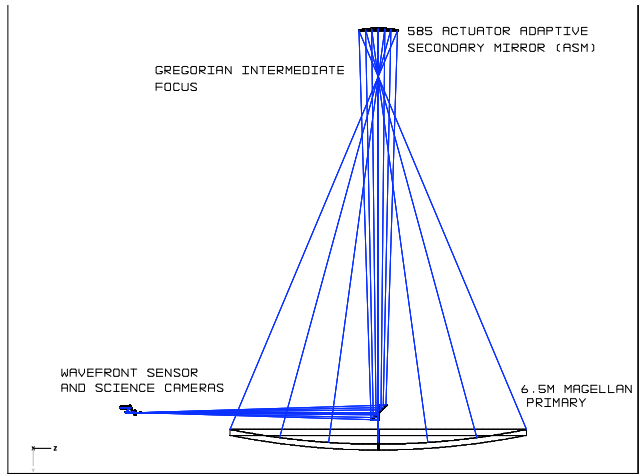
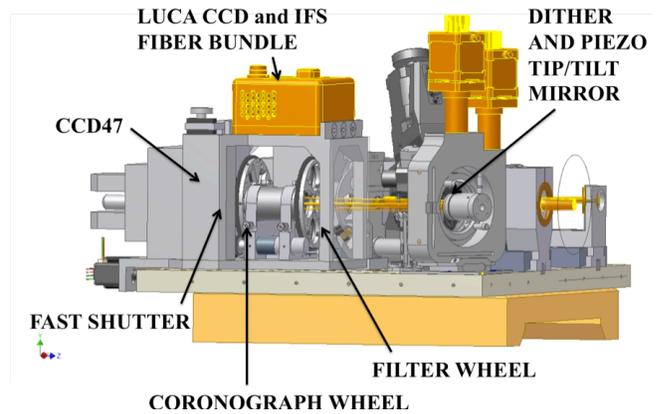

**Figure 1:** Raytrace of the 6.5 m Magellan Telescope with the F/16 adaptive secondary. Note the concave secondary and the intermediate Gregorian focus. Our science instruments are located at the nasmyth focus.

**Figure 2a:** Side view of the VisAO arm of the W-unit. A beamsplitter after the filter wheel sends light up to a port containing either the LUCA tip/tilt CCD or the IFS fiber bundle.

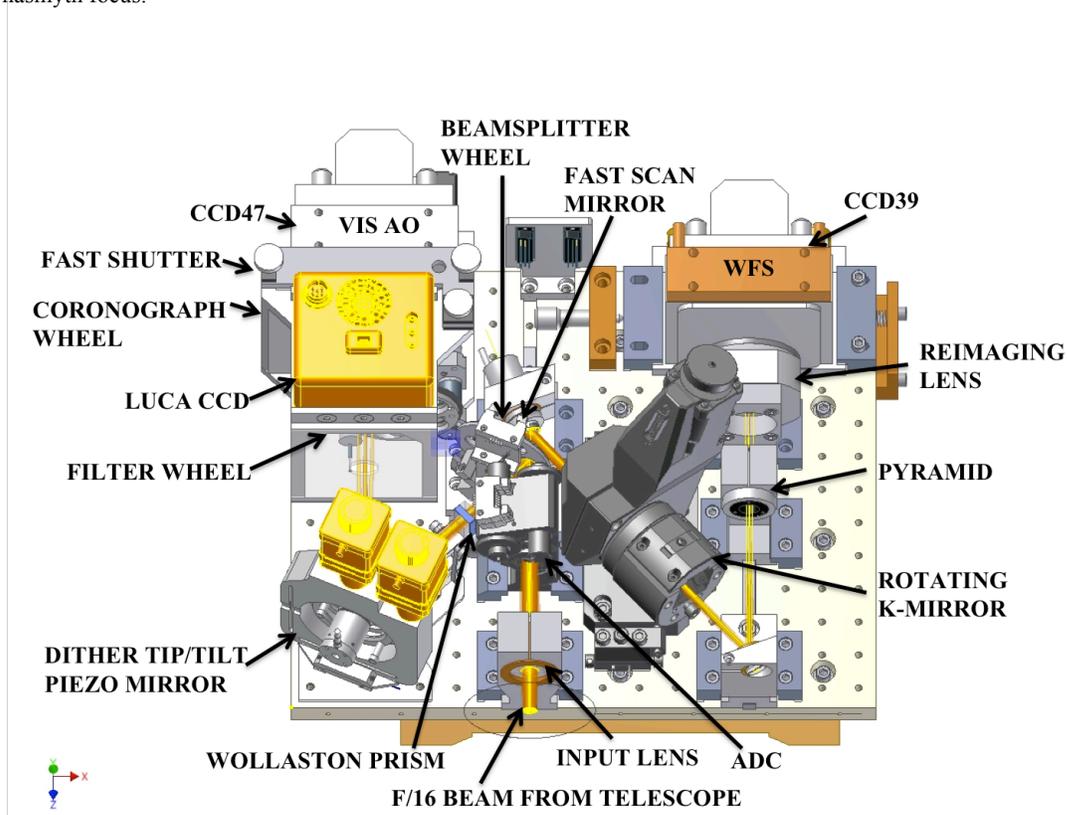

**Figure 2b:** The W-unit layout. Light from the Magellan F/16 focus enters through the input lens and then passes through the ADC before hitting a beamsplitter wheel. Transmitted light is sent to the pyramid wavefront sensor on the right side of the board. Reflected light is sent to the VisAO CCD and IFS on the left side of the board.

FOV will typically be limited by the isoplanatic angle to less than 8.5"). If we estimate no better control than these current systems, and note that our fitting error is a factor of 2x rad$^2$ better, then it is clear that our Strehls with bright stars (fitting error limited) will trend towards I band Strehls of 16%. Our objective for the CCD47 is diffraction limited image quality over the full 8.6" FOV over the band 0.5-1.0 μm out to a Zenith angle of 70°. Magellan is 2.7 times larger than HST and therefore has a diffraction limit 2.7x sharper at the same λ. However, all the existing operational cameras on HST do not Nyquist sample wavelengths less than 1μm (it should make a ~48mas FWHM image at 0.6μm). Therefore, our VisAO camera will make images >4.7x better than HST in terms of pixel resolution.

The VisAO camera will have the option to either make visible AO images with the CCD47 and 8.5 mas pixels, or to take spatially resolved spectra in IFS mode with 20 mas fiber elements. In either case, IR science can simultaneously be performed with the BLINC/MIRAC4 instrument. The fine spatial and spectral (R~1,800) resolution of the fiber-fed IFS will open a window to a panoply of science, such as studying emission lines in the inner regions of Herbig Ae/Be disks, resolving tight astrometric binaries, mapping the surface of Titan, etc. For a more detailed discussion of the science capabilities and motivations for the VisAO IFS, see Follette et al, these proceedings.

## 2. THE W-UNIT: PYRAMID WAVEFRONT SENSOR, VISAO SCIENCE CAMERA, AND FIBER-FED IFS

The W-Unit (Figures 2a and b) is an optical board located on a three-axis translation stage that can patrol a 2.3x3.2 arcmin field at the Nasmyth focal plane in order to acquire NGS guide stars and VisAO science targets. The W-Unit contains two optical channels: the pyramid wavefront sensor channel and the VisAO science channel. The VisAO science channel can be configured to work either in imaging mode with the CCD47 or in IFS mode with the fiber array. Incoming visible light passes through a telecentric lens and a triplet lens that converts it from a diverging F/16 beam into a converging F/49 beam. This light then passes through the ADC before hitting a beam splitter wheel. Light transmitted by the beamsplitter goes to the WFS and reflected light goes to the VisAO channel.

**2.1 Pyramid Wavefront Sensor Channel**

The WFS channel consists of a fast steering mirror, a K-mirror rerotator, a double pyramid, a reimaging lens, and the CCD39. The resultant image on the CCD39 is four pupil images whose intensity variations can be used to reconstruct the wavefront. A detailed description of the operation of the pyramid sensor (PS) arm of the W-unit can be found in Esposito et al. 2008. The pyramid sensor is very important for visible AO because of its potential for diffraction limited performance and variable sensitivity. A Shack-Hartmann (SH) sensor is diffraction limited by the size of a pupil sub-aperture: i.e. the pitch of the lenslet array. The PS uses the full pupil aperture and is only diffraction limited by the size of the primary mirror. Since the wavefront sensing wavelengths are essentially the same as the science wavelengths (~0.7 μm), it is essential that the wavefront sensor be as close to the diffraction limit as possible (Esposito et al. 2000). The dynamic range provided by the PS should allow us to use ~2 mag fainter guide stars than would be allowed by the SH sensor, in addition to allowing us to come very close to diffraction limited correction (Esposito et al. 2001).

**2.2 The VisAO Optical Path**

The light reflected from the beamsplitter will travel to the VisAO camera, which has two modes: an imaging/acquisition mode that uses an E2V CCD47 and an IFS mode that uses our custom fiber bundle to bring light to the LDSS3 facility spectrograph. Light reflecting off of the beamsplitter wheel first passes a stage that holds either the Wollaston prism or the first IFS triplet lens and can move either in or out of the beam. It then reflects off of a λ/20 silver mirror that is mounted to both a fast piezo for AO tip/tilt correction and a standard actuated tip/tilt stage for beam steering and dithering. The light then passes through the filter wheel before hitting a fixed 45° glass plate that has a chrome coronographic spot in its center to act as both an anti-blooming device for the CCD47 and as a pick-off mirror sending light from our bright guide star to the LUCA CCD for tip/tilt sensing, when in imaging mode. Light transmitted through the glass plate passes through a wheel holding a variety of coronographic spots of various sizes and apodizations, in addition to our custom SDI filters (see Table 1). Finally, the light passes through our fast asynchronous shutter before hitting the CCD47. The converging F/49 beam results in a square FOV of 8.6". Our goal for the CCD47 is diffraction limited image quality over the full 8.6" FOV over the band 0.5-1.0μm out to a Zenith angle of 70°. In IFS mode, the 45° glass plate is replaced by a beamsplitter that reflects light up to our fiber array and transmits a small amount of light to the CCD47, which now acts as the tip/tilt sensor.

| Position | Beamsplitter | Filter Wheel | Coronagraph |
|---|---|---|---|
| 1 | Bare glass (96/4) | SDSS z' | Ha SDI |
| 2 | 50/50 | SDSS i' | [OI] SDI |
| 3 | Dichroic (reflect l <800 nm) | SDSS r' | [SII} SDI |
| 4 | Dichroic w/ ND 1 filter | open | 3" ND 1 |
| 5 | Metallic ND 3 | Long-pass "y" (l > 950 nm) | 1" ND 1 |
| 6 | ----------------- | --------------- | open |

**Table 1:** VisAO components

### 2.3 W-Unit Components: Wollaston SDI, Filters, Beamsplitters, Coronagraphic Spots

The W-unit has several wheels and stages that hold many components to enable a large and flexible suit of science modes and measurements. Because of our narrow FOV and our bright guide star requirement, the science cases are necessarily constrained to circumstellar science: exoplanets, disks, etc.

Our pupil is reimaged near the beamsplitter of the W-unit. Depending on the science case and the brightness of the guide star, a beamsplitter can be selected to steer various proportions of the light to the WFS and the VisAO science arm. Or, a dichroic can be selected that will steer blue photons to the VisAO arm and red photons to the WFS. In the VisAO arm, there is a filter wheel that holds various standard band-pass optical filters: SDSS r', i', and z', along with a long pass filter ($\lambda > 950$ nm), and an open position. Because the responsivity of our CCD47 falls off around 1.05 μm, our long pass filter effectively acts as a y filter.

The VisAO imaging mode will also have SDI imaging capability: a powerful technique for mitigating speckles and noise when looking at spectral lines (see Close et al. 2005). When in SDI mode, a Wollaston prism designed for an angular deviation of 1.15° can be placed in the beam approximately 60 mm after the pupil. This prism splits the beam into two orthogonally polarized beams that are imaged onto different halves of the CCD47. In front of the CCD is a second wheel (which we call the coronagraph wheel) that contains coronagraphic spots and special SDI filters. The SDI filters are actually two rectangular filters placed next to each other, each of which covers half of the CCD, corresponding to one of the two images coming from the Wollaston prism. One of the rectangular filters is a narrow band filter centered on a given spectral line and the other is center at the continuum, slightly off of the line. Subtracting the continuum image from the spectral line image results in an image of the spectral line structures only (disk, jets, companions, etc.), almost entirely free from speckles and other broadband sources of noise. In this fashion, we are currently planning to be able to perform SDI at Hα, [OI], and [SII]. See Follette et al for more discussion of the science motivations behind these spectral lines.

### 2.4 Tip/Tilt Loop

Our VisAO optical path includes a tip/tilt AO loop that will mitigate jitter and enable us to achieve our ~20% predicted Strehl (see Males et al, these proceedings). The pick-off spot sends light to the LUCA CCD, which acts as the tip/tilt sensor. The LUCA CCD controls our fast tip/tilt mirror and is able to operate at 2KHz. Because this loop is so late in the optical train, the amount of non-common path tip/tilt error should be extremely low.

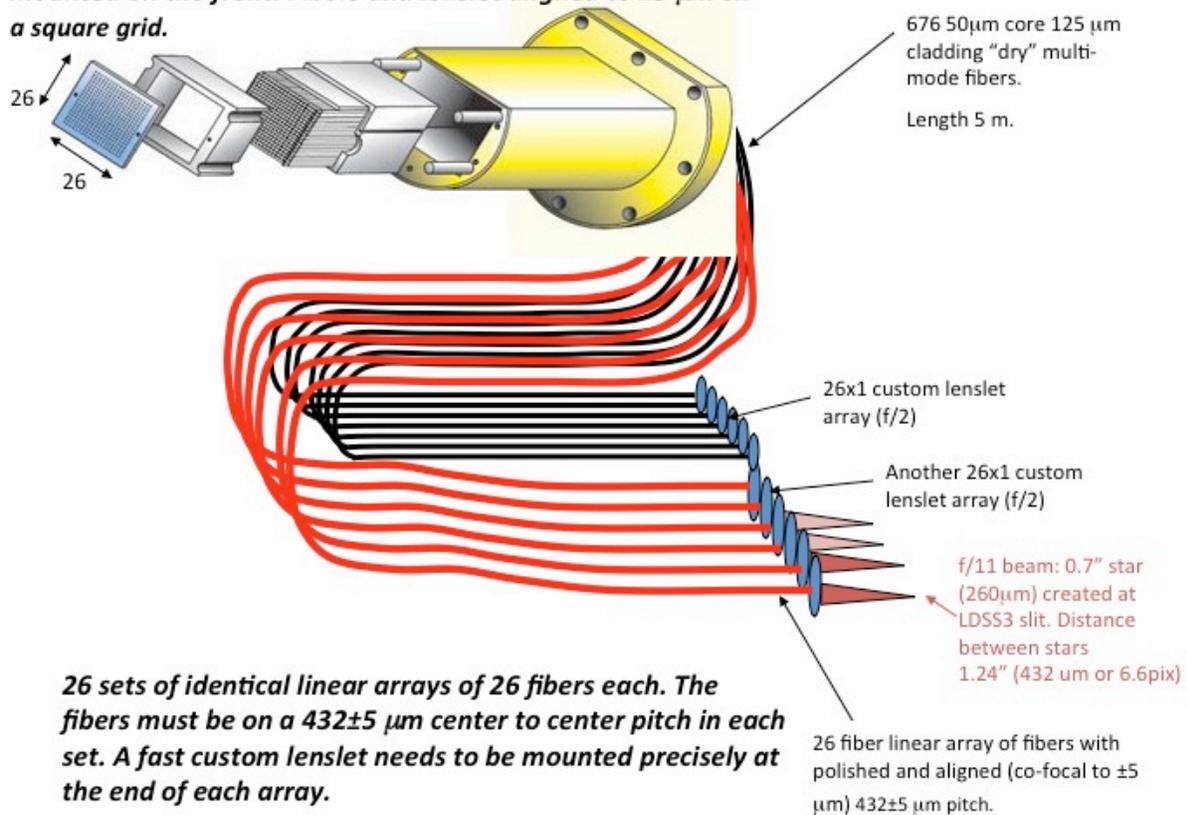

**Figure 3:** IFS Schematic. A 26x26 array of fibers takes light from the VisAO focal plane to the LDSS3 spectrograph where the fibers are coupled to the wide-field LDSS3 slit.

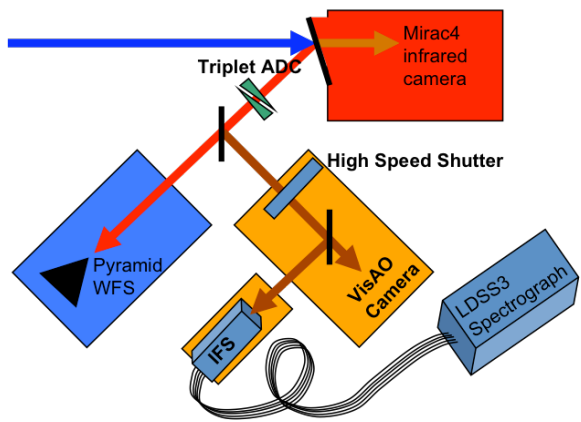

**Figure 4:** IFS and VisAO block layout. AO corrected light is split by a dichroic and a beamsplitter between the MIRAC4 IR science camera, the pyramid WFS, and the VisAO science instrument that has the option of sending light to the fiber-fed IFS.

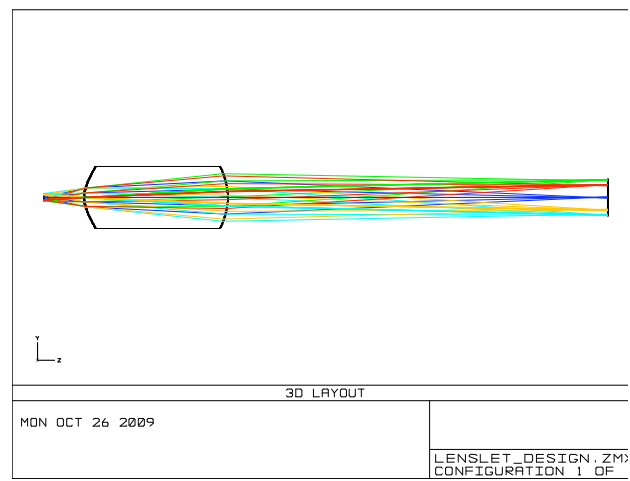

**Figure 5:** Our custom aspheric output lenslet design that will couple the ~F/2.5 output of the IFS fibers to the F/11 LDSS3 slit.

# 3. THE INTEGRAL FIELD SPECTROGRAPH

In addition to the VisAO CCD47 imaging mode, we have designed an IFS that will be able to take full advantage of our 4-8 arcmin AO corrected isoplanatic patch. A removable beamsplitter will direct light towards a custom 26x26 array of optical fibers spaced with 160 μm pitch (Figure 3). A microlenslet array positioned in front of the bundle will improve our fiber coupling efficiency and give us a fill-factor of ~99%. These fibers will transmit 0.6-1.05 μm light from the AO corrected focal plane to LDSS3, a wide-field red-sensitive facility spectrograph (Figure 4).

The IFS mode will have two different plate scales: 20 mas/pixel and 105 mas/pixel. In order to slow the beam down enough to transition from 105 mas/pix to 20 mas/pix (F/49 to F/225) we have designed two custom triplet lenses that will both swing into the beam when fine 20 mas spectral imaging is desired (see Figure 6). The first triplet will be interchangeable with and in the same location as the Wollaston prism. The second triplet will be captured inside the baffle tube that holds the beamsplitter that folds the beam up to the fiber bundle. These two optics acting together will allow us sufficient optical leverage to slow the beam down enough to reach our desired plate scale while still keeping optical and chromatic aberrations below the diffraction limit. For the raytrace and spot diagrams of the two-triplet fine plate scale optical design, see Figures 6 and 7. The 20 mas/pix mode will be the highest resolution visible IFS instrument in the world. The coarser 105 mas/pixel plate scale will be used as a "faint object" mode for science targets too faint to be detected in the 20 mas/pixel mode.

Another optical design challenge presented by the VisAO IFS is the coupling of the output end of the fibers with the input slit of LDSS3. Typical high numerical aperture fibers have an output f-ratio of ~F/2-3. The LDSS3 slit has an input focal ratio of F/11. In order to couple this fast fiber output with the slower slit input, we have designed a two-sided aspheric lenslet array. This lenslet array is made of fused silica and is within reasonable lithographic fabrication constraints, which are chiefly governed by the maximum sag of the lens profile. The raytrace of an element of our custom aspheric lenslet array is shown in Figure 5.

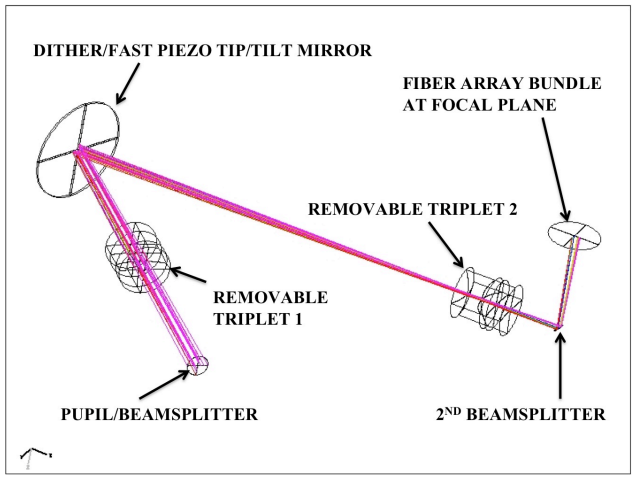

**Figure 6:** Raytrace of the 20 mas mode of the VisAO IFS from the W-unit beamsplitter to the IFS fiber focal plane. The first and second triplets can move in or out of the beam to switch between the F/225 20 mas mode and the F/49 105 mas mode.

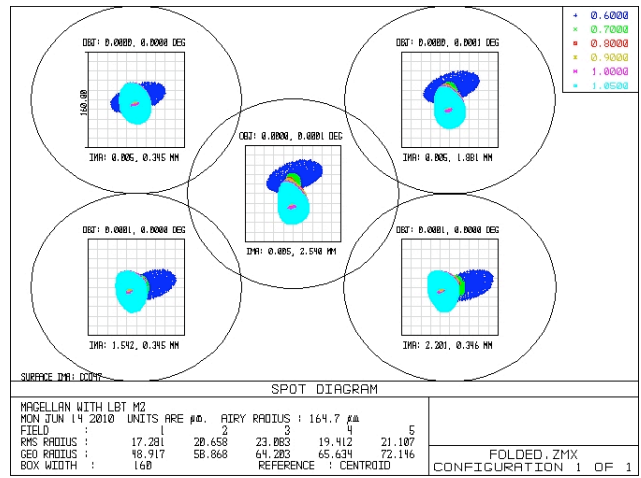

**Figure 7:** Spot diagrams of the VisAO focal plane (fiber entrance) in 20 mas mode. The circles are the first airy minimum at 600 nm and the squares are the size of a lens element (160 μm x 160 μm).

# 4. THE ADVANCED TRIPLET ADC LAB RESULTS

Achieving diffraction limited performance over our broad visible science band requires that 2000μm of lateral color be corrected to better than 10μm. Traditional atmospheric dispersion correctors (ADCs) consist of two identical counter-rotating cemented doublet prisms that correct primary chromatic aberration. Our novel 2-triplet ADC design uses two counter rotating cemented triplet prisms made of both normal and anomalous dispersion glasses to correct both primary and secondary chromatic aberrations. At high Zenith angles, this design is predicted to perform 58% better than the traditional two-doublet design. The criteria used to evaluate relative performance of various designs is the total rms spot size relative to the spot centroid for six different wavelengths spanning the 0.5-1.0μm range in increments of 0.1μm. The Zemax "Atmospheric" surface was used to simulate the atmospheric dispersion with estimated Magellan site parameters (humidity, temperature, etc.). The ADC also serves to increase the sensitivity of the PS by allowing smaller scan modulations due to the smaller spot size at the pyramid tip focal plane. For details of our advanced ADC design and additional analysis, see Kopon et al, 2008 and Kopon et al, 2009.

Because we cannot fit a 6.5m telescope and an atmosphere in our lab, we have devised a white-light point source test for our newly fabricated ADCs that precisely measures the chromatic dispersion of the ADC.

## 3.1 The 2-Triplet Design

Most ADCs designed and built to date consist of two identical counter-rotating prism doublets (often referred to as Amici prisms) made of a crown and flint glass. The d-light (587 nm) indices of the two glasses are matched as closely as possible in order to avoid steering the beam away from its incident direction. The wedge angles and glasses of the prisms are chosen to correct primary chromatic aberration at the most extreme zenith angle. By then rotating the two doublets relative to each other, an arbitrary amount of first-order chromatic aberration can be added to the beam to exactly cancel the dispersive effects of the atmosphere at a given zenith angle. The 2-Doublet design corrects the atmospheric dispersion so that the longest and shortest wavelengths overlap each other, thereby correcting the primary chromatism. Secondary chromatism is not corrected and is the dominant source of error at higher zenith angles (Figure 9). To correct higher orders of chromatism, more glasses and thereby more degrees of freedom are needed.

In our 2-triplet design (Figure 8), a third glass with anomalous dispersion characteristics (Schott's N-KZFS4) is added to the crown/flint pair. Like the doublet, the index of the anomalous dispersion glass was matched as closely as possible to that of the crown and flint. The Zemax atmospheric surface was set to 70 deg zenith and the relative angles of the ADC were set to 180 deg. The wedge angles of the three prisms in the triplet were then optimized to correct both primary and secondary chromatism.

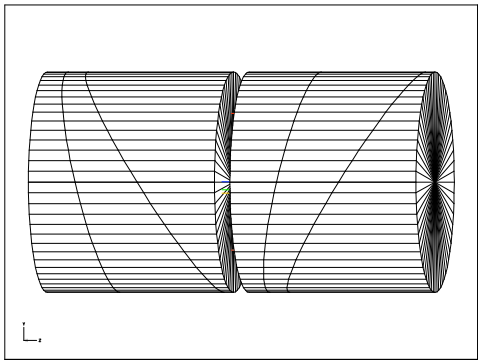

**Figure 8:** The advanced triplet ADC. The two counter-rotating triplets correct both primary and secondary color out to 70° zenith angle.

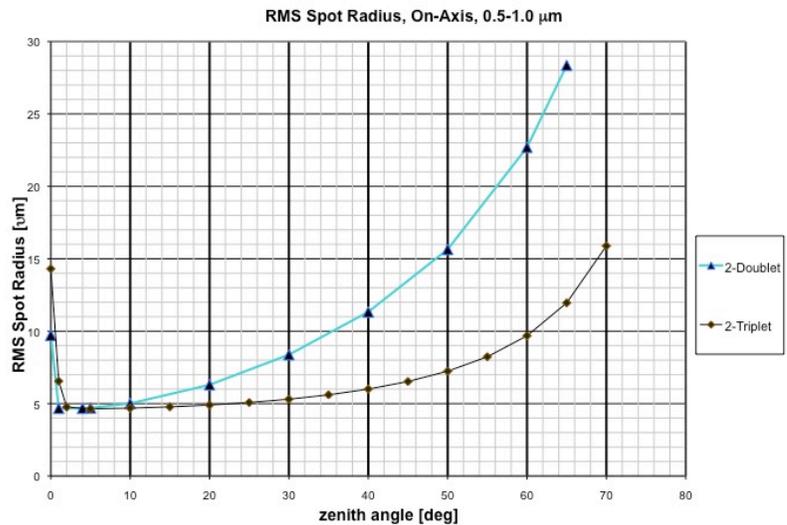

**Figure 9:** RMS spot size as a function of zenith angle showing the performance of a conventional ADC design and our novel 2-triplet design. Our design performs 58% better at higher zenith angles.

## 3.2 The Lab Test and Results

When used on-sky looking at a broadband point source, such as a star, the ADC will be taking a little linear rainbow of light, which has been dispersed by the atmosphere, from the Magellan telescope focal plane and correcting it so that it falls on the CCD47 as a well-corrected broadband point. However, in our lab we cannot easily and reliably simulate the dispersion effects of the atmosphere in order to generate that "rainbow". Therefore, we have designed a test that works in reverse: the ADC takes a white-light point source and disperses it into a rainbow at the focal plane (Figure 11). Using narrow band filters, we measure where various different wavelengths fall on the focal plane and compare these locations to the Zemax predictions.

Using a microscope objective and a pair of achromatic doublets, we reimage a fiber white-light source onto a 10μm pinhole (Figure 10a). This pinhole serves as the point source that is located where the nominal Magellan telescope F/16 focal plane would be. The point source feeds the W-unit triplet input lens, which converts the beam from a diverging F/16 beam to a converging F/49 beam. This converging white light beam then passes through the ADC triplet prisms, which are in rotating mounts. The image is then measured with our Electrim EDC-3000D lab CCD. By placing three different narrow band filters in front of the white light source (532 nm, 850 nm, and 905 nm) and measuring where these wavelengths fall on the focal plane relative to each other for various relative ADC clockings, we are able to measure both the primary and secondary dispersion characteristics of the ADC.

The results of this test and the Zemax theoretical predicted curves are shown in Figure 12. The most accurate results, as expected, are given by the largest ADC relative clocking angle and the largest wavelength difference. In these cases, the ADC performance differs from the Zemax prediction by ~0.3-0.6%, which is on the order of the estimated error of our lab setup. Therefore, the ADC dispersion is correct to within our measurement error, confirming that the ADCs were fabricated correctly, are chromatically well behaved, and perform as expected.

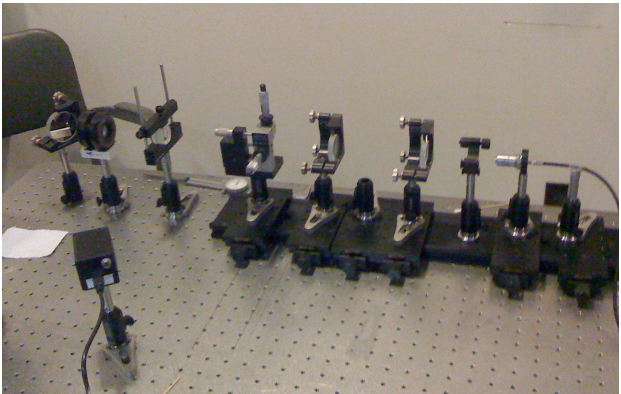

**Figure 10a:** Our ADC lab test setup. A white light fiber source feeds a microscope objective and is reimaged onto a 10 μm pinhole which simulates the F/16 focus of the Magellan telescope.

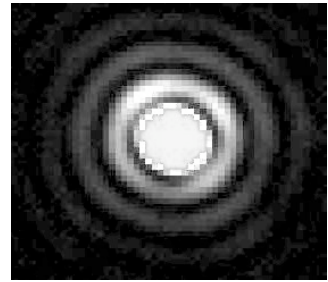

**Figure 10c:** The 0.96 Strehl Airy pattern produced by our VisAO optics, including the ADC, at λ=531 nm.

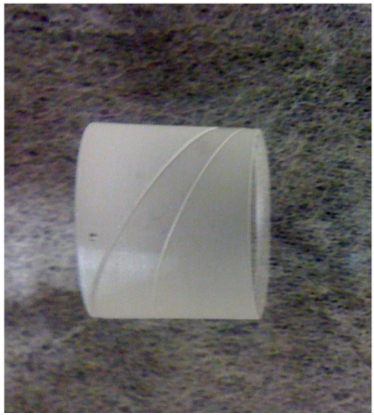

**Figure 10b:** Photo of the as fabricated ADC.

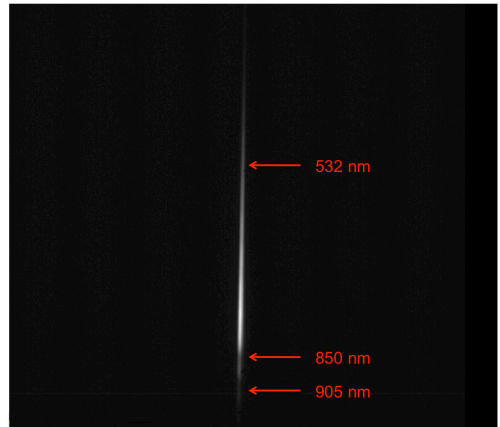

**Figure 11:** The white light point source after being dispersed into a line by the ADC. Using narrow band filters at 532, 850, and 905 nm, we measured the relative displacements of these wavelengths at the focal plane and compared them to the Zemax predictions (see Figure 12).

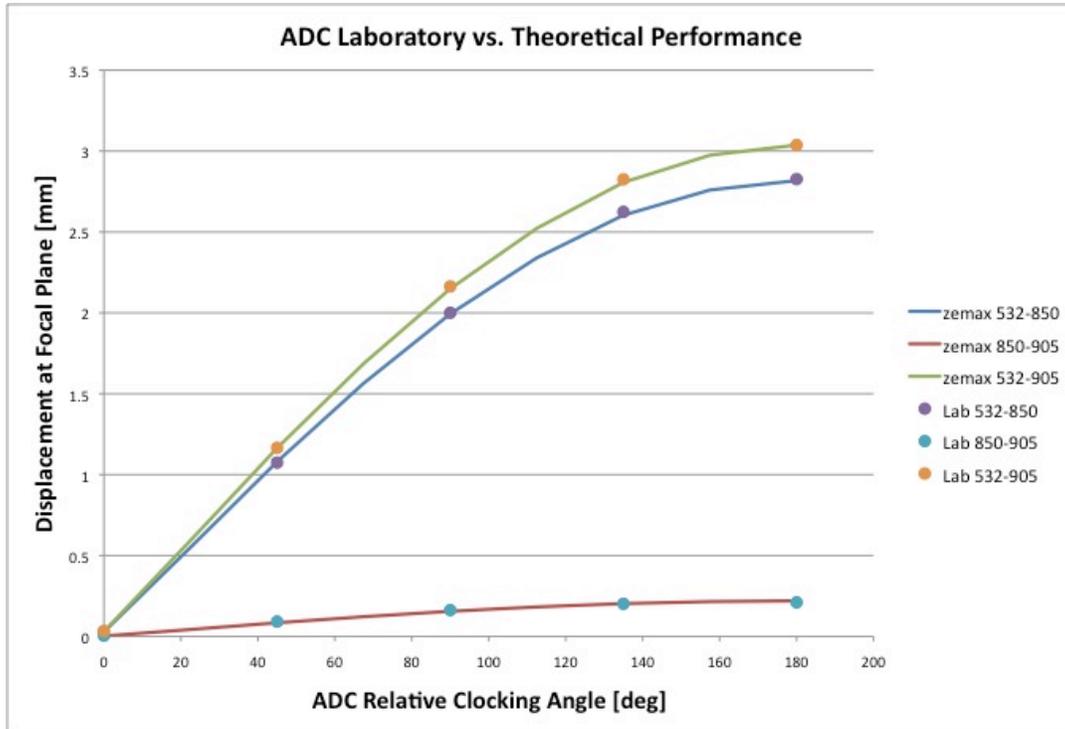

**Figure 12:** Plot comparing the measured and theoretical displacements of individual wavelengths at the focal plane. The x-axis is the relative clocking of the two ADC triplets relative to each other. The y-axis is the displacement in mm between two wavelengths. The error bars are smaller than the spots used on the plot. This test demonstrates that the chromatic dispersion of our as-fabricated ADCs is the same as that predicted in our Zemax model.

## 5. THE CUSTOM WIDE-FIELD ACQUISITION AND ACTIVE OPTICS CAMERA

Each non-IMACS Magellan instrument has its own guider assembly(s) that serves as both a wide-field acquisition and guiding camera and active optics low order wave front sensor (Schechter et al. 2002). The guider assemblies have pneumatic slides that switch back and forth between two sets of optics: wide-field imaging optics for acquisition, guiding, and seeing estimates; and a Shack-Hartmann mode with a collimating double lens that reimages the pupil onto a lenslet array in front of the CCD. Most of the Magellan guiders currently on the mountain make use of off-the-shelf camera lenses for their wide-field imaging. The optical quality of these lenses limits the image quality of the guide camera to 0.4 arcsec FWHM. Since this is of the same magnitude as the seeing, these cameras cannot effectively measure the seeing on good nights.

To overcome this limitation, we have designed a new custom 50" FOV acquisition camera that can be placed in the center of the field (Figure 14). The design residual of the new wide-field lens is significantly better than seeing limited (0.1") over its 50" field. The wide-field lens will be in a tube that is on a mechanical slide that can move in front of the CCD when wide-field acquisition mode is desired. The lens group operates at F/8.25 (giving 0.05"/pix on each 13μm pixel, but really 0.10"/pix since 2x2 binning is standard) and is well-corrected over the band 550-850 nm. The end of the tube holds a filter that limits the transmitted light to this spectral range. The pneumatic slides allow us to switch easily from the wide-field lens to the SH mode and back. The spot diagram in Fig. 14 shows the quality of the wide field camera design. The black square is the size of one of the Magellan 1k x 1k E2V CCD pixels (0.1" x 0.1" when used in standard 2x2 binning). This design residual is far better than even the best (~0.25") optical seeing conditions that are possible at the telescope. We have also changed the focal length of the collimating lens of the Shack-Hartmann E2V CCD to accept the F/16 beam of our secondary mirror, instead of the standard F/11 beam. The layout of our guider ring with the positions of the W-unit and the guider probe is shown in Figure 12.

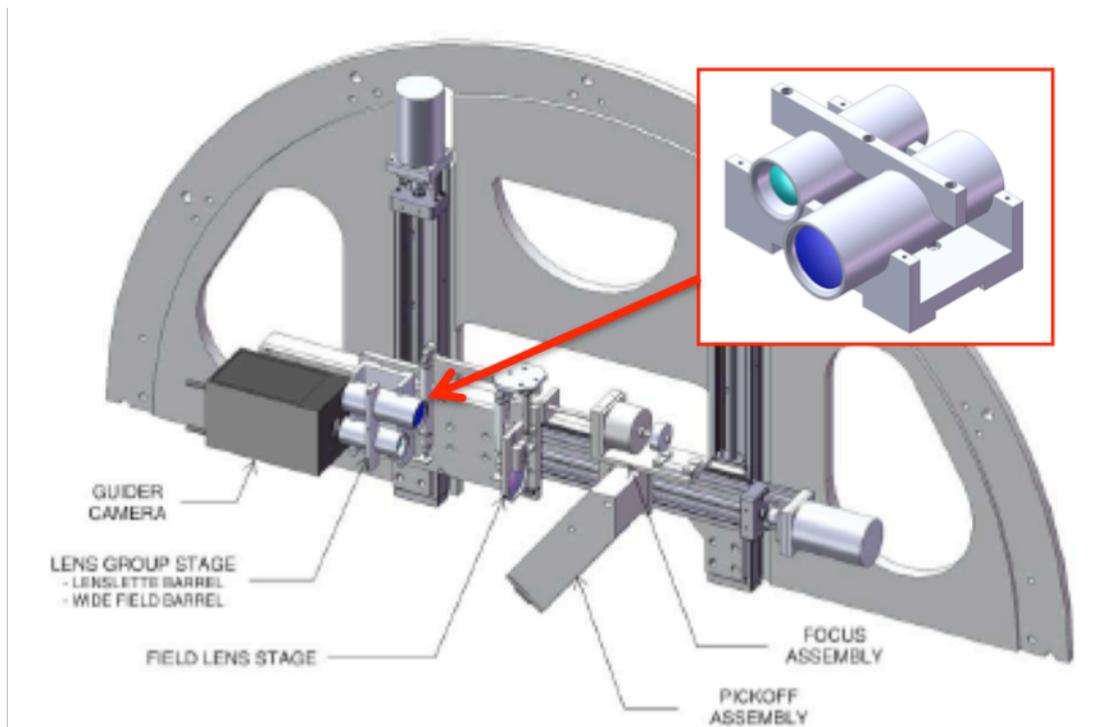

**Figure 13:** The guider assembly. The pickoff mirror sends light from a star within the patrol field to the focus assembly and then the lens group stage where the beam passes through either our custom wide-field lens or a collimating lens followed by a lenslet array for Shack-Hartmann wave front sensing. Both modes use the same 1k x 1k E2V CCD.

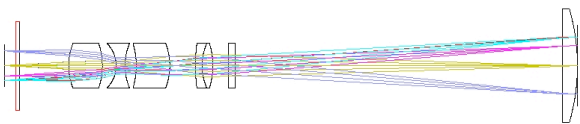

**Figure 14:** Ray trace of our 50" FOV acquisition camera. Light enters from the right through the field lens, which is on a separate mechanical stage at the F/16 Magellan focus. The smaller lenses are in a lens tube that can be moved in and out of the beam to switch between the wide-field imaging mode and the Shack-Hartmann mode.

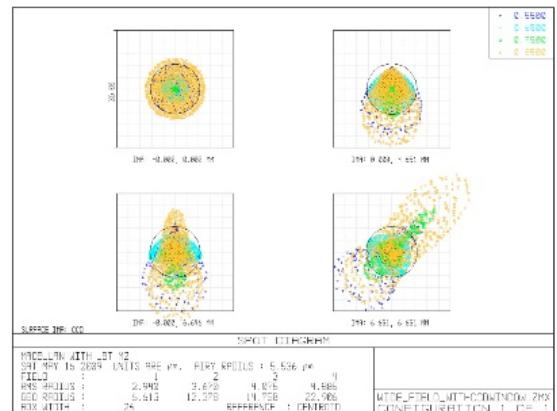

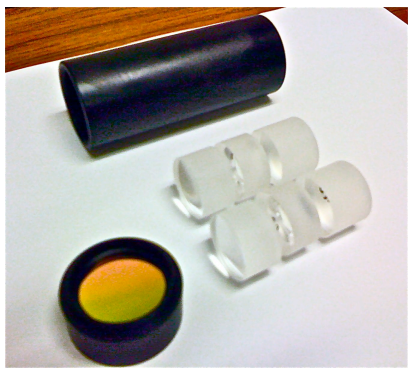

**Figure 15:** The custom wide-field singlet lenses, filter, and lens tube.

**Figure 16:** Spot diagrams for the wide-field lens over the band 550-850 nm. From top left to bottom right: on-axis, 7/10 field point, CCD edge (25" from on-axis), and CCD corner. The black circle is the 550 nm diffraction limit. The black square is 1/10 of an arcsec on a side.

## 6. CONCLUSION

In this paper we have presented the design of our VisAO visible adaptive optics instrument, which is part of the Magellan AO system. The VisAO camera will have the ability to operate simultaneously with the MIRAC4 10 μm IR camera. The VisAO camera will have two modes: the CCD47 imaging mode with 8 mas pixels and the fiber array IFS mode, which has plates scales of 20 mas and 105 mas per element. We also present the lab results of our recently fabricated triplet ADC and the design of our custom wide field guider camera. The Magellan AO system is scheduled for first light in early 2012.

## ACKNOWLEDGEMENTS

This project owes a debt of gratitude to our partners and collaborators. The ASM and WFS could not have been possible without the design work of Microgate and ADS in Italy as well as Arcetri Observatory and the LBT observatory. We would like to thank the NSF MRI and TSIP programs for generous support of this project in addition to the Magellan observatory staff and the Carnegie Institute. The first author would like to thank SPIE for its continued support through the SPIE Scholarship Program. We would also like to thank the engineers and staff at Optimax for doing an excellent job fabricating the ADC, a challenging and unique optic.